\documentstyle[aps,floats,pre,preprint,tighten,epsf]{revtex}
\draft
\begin{document}
\title{The selection of altruistic behaviour}
\author{R.Donato$^1$\thanks{E-mail: {\tt donato@roma1.infn.it.}}, 
L.Peliti$^2$\thanks{Associato INFN, Sezione di Napoli.
E-mail: {\tt peliti@na.infn.it.}}, M.Serva$^3$\thanks{To whom correspondence
should be addressed. Tel.: +39-6-4991-3840.
Fax: +39-6-446-3158. E-mail: {\tt serva@roma1.infn.it.}}}
\address{$^1$ Dipartimento di Fisica, Universit\`a di Roma
``La Sapienza'' \\P.le Aldo Moro 2, I-00185 Roma (Italy)\\
$^2$ Dipartimento di Scienze Fisiche and Unit\`a INFM,\\
Universit\`a di Napoli ``Federico II''\\
Mostra d'Oltremare, Pad.~19, I-80125 Napoli (Italy)\\
$^3$ Dipartimento di Matematica, Universit\`a dell'Aquila and Istituto
Nazionale di Fisica della Materia\\
I-67010 Coppito (AQ) (Italy)}
\date{\today}
\maketitle
\begin{abstract}
Altruistic behaviour is disadvantageous for the individual while is 
advantageous for its group.
If the target of the selection is the individual,
one would expect the selection process to lead to populations formed
by wholly homogeneous groups, 
made up of either altruistic or egoistic individuals,
where the winning choice depends on the balance beetwen group advantage and 
individual disadvantage.
We show in a simple model that populations formed
by {\em inhomogeneous\/} groups can be 
stabilized in some circumstances.
We argue that this condition is 
realized when there is a relative advantage conferred by the 
presence of a few altruists to all the members of the group.
\end{abstract}


\leftline{Keywords: deme, quasispecies, kin selection, group selection,
phase transition}

\section{Introduction}
Natural selection is an egoistic process: individual organisms compete for 
representation of their genes in the next generation, and only those genomes 
better able to survive and reproduce are likely to be maintained in the 
evolutionary process. In this framework it is difficult to explain how
``altruistic'' genes (that is genes which determine a behaviour 
disadvantageous to the carrier, but beneficial 
for other individuals) can be selected.

Nevertheless, we can find in nature some examples of
altruistic behaviour: social organization in insect societies
\cite{Ham1,Ham2}, parental care \cite{MS1,MS3},
warning calls in birds \cite{MS2}, and so on. 
So far, there is no evidence of a genetical basis for altruistic behaviour.
However, it is an interesting problem in itself to understand how interactions 
between individuals may be reflected on the action of individual natural 
selection. 
The presence of altruists increases the chances of the group which 
it belongs to. The general problem is not only to understand what is the 
level of selection action, but also how the different levels may interact under 
the selective pressure.

Classically, two main mechanisms have been proposed to explain the origin and 
evolution of altruism: kin selection \cite{Hal}, \cite{Ham1,Ham2}, 
\cite{MS1,MS3}, and 
group selection \cite{WE}, \cite{Esh}.
In both mechanisms it is necessary to assume that individual fitness depends
not only on individual genotype, 
but also on the genetical composition of the population.
In kin selection the 
individual probability of survival is proportional to the relatedness that the 
individual shares with the altruist. The selection mechanism thus
is limited only to related individuals, and the selection unit is
the individual.
In group selection, the interaction takes place 
between individuals belonging to the same group, and groups are defined by 
external constraints (for example spatial boundaries), disregarding any 
relatedness between individuals. 
In most cases, one ends up with a situation in which either
egoistic or altruistic behavior is selected. Nevertheless, it is clear from
observation that most often altruistic behavior is exhibited only by
some individuals in the group.

A recent report concerns territorial defense by prides of lions.
Female lions living in Serengeti National Park and Ngorongoro Crater (Tanzania)
live in groups that hunt togheter and defend a common territory against other 
groups. Heinsohn and Packer report in \cite{sci} that some lionesses act
more aggressively and swiftly in defense, while others tend to lag behind.
No consistent 
relations were observed between physical size and defense-readiness. 
Furthermore, when engaging in aid of a pride-mate mounting a defense, no 
consistent relation was found between the readiness of the helper, and the kin 
relation between the helper and the first defender.
Territorial defense in groups of female lions therefore seems to be a case of 
altruistic behaviour where kin relation does not play a role.

We show in the following, in an exactly solvable
model, that altruistic behavior
can be selected both disregarding relatedness between individuals
and maintaining the individual as the selection unit: morevoer, that, if
the advantage of the presence of altruistically behaving
individuals is already felt when their number is small,
the stable situation is one in which {\em inhomogeneous\/} groups,
formed of both egoistic and altruistic individuals, coexist
with both fully egoistic and fully altruistic individuals.

\section{The model}
We consider a population made up of $M$ individuals, divided in
$N$ groups (demes) of $L=M/N$ individuals on average.
We assume non-overlapping generations,
heredity acting according to the usual Mendelian mechanism, and the
presence of a behavioral locus with two alleles: A (recessive)  and 
E (dominant).
An individual of genotype AA is altruist, while AE and EE are not.
However, as a first approximation,
we disregard the individual genetical structure, considering only the two 
classes of individuals: altruist (A) and egoist (E).
Demes have a variable number $j$ of altruistic individuals A.
If, for a given deme, $j$ is greater than a threshold $j^{\ast}$ the deme
is called {\em altruistic}, otherwise {\em egoistic}. 
We shall call any deme with exactly $j$ altruistic individuals
a $j$-deme.

To each individual $\alpha$ in a generation we associate its fitness value
$f(\alpha)$, which is proportional to the survival 
probability of its genome in the population over one step. 
The fitness $f(\alpha)$ of the individual $\alpha$ depends on two factors:
\begin{enumerate}
\item {\em on its genotype\/}: 
if the individual is altruist (genotype A), its fitness
is {\em reduced\/} by a factor $(1-r)$ with respect to the other members of its
deme;
\item {\em on its deme composition\/}:
if the individual (whatever its genotype) belongs to an altruistic deme
its fitness is {\em enhanced\/}  by a factor $1/(1-c)$, with respect
to individuals with the same genotype belonging to an egoistic one.
\end{enumerate} 
The parameter $r$ is called {\em intrademic selection rate\/} and $c$ is
called {\em interdemic selection rate}.
This form of the fitness is summarized in Table 1.

\begin{table}
\begin{tabular}{c|cc}
 \phantom{$j<J^{\ast}$}      & $j < j^{\ast}$       &$j\geq j^{\ast}$ \\ 
                                                      \hline
A& $1-r$                &$(1-r)/(1-c)$ \\ 
E& $1$                  &$1/(1-c)$     \\   
\end{tabular}
\caption{Fitness table. Rows: {\em individual\/} genotypes; columns: genetical
{\em group\/} composition.\label{tabfit}}
\end{table}

The population evolves according to a discrete process, involving the 
following steps:
\begin{enumerate}
\item {\em Selection and reproduction\/}:
Taking into account the internal genetical structure, one can consider both
asexual and sexual reproduction mechanisms.
In the following, we will consider only the asexual case, but the reproduction
mechanism does not affect the essential results.
The average number of the offspring of an individual $\alpha$ is given by
\begin{equation}\label{pripasex}
\overline{\nu(\alpha)}=\frac{M\,f(\alpha)}{\sum_{\beta=1}^{M}f(\beta)}\ ,
\end{equation}
where $f(\alpha)$ is the fitness of the individual $\alpha$,
as defined in Table~\ref{tabfit}, and the sum runs 
over the whole population. At the end of this step, the offspring of the individuals of one deme in one
generation form one deme.
\item {\em Deme splitting\/}: The mean size of the demes is equal to $L=M/N$.
However, after the reproduction step, some demes can disappear
and the mean size of the surviving demes
would correspondingly increase. In particular, some demes may have become 
much greater than $L$. We assume that too large demes split in such a way
as to keep the mean size $L$ constant.
This rule is motivated by the analogy with social animals, which do not
typically live in too large groups. It is also a necessary 
ingredient for purely modelling reasons: if there is no maximal size of demes,
sooner or later one of the demes would take over the whole population, and 
interdemic selection would cease to operate.
\end{enumerate}
Notice that an A individual has always less 
chances than an E individual belonging to
the same deme, but all the individuals
in an altruistic deme may have a greater survival probability than individuals
belonging to egoistic demes, independently of their genotypes.

\section{The master equation}
We assume that the total number of individuals $M$
tends to infinity along with deme number $N$,
keeping the typical deme size $L$ finite.
%Define $N=M/L$.
We denote by $n_j(t)$ the number of $j$-demes
at generation $t$, and by $\rho_j(t)=n_j(t)/N$
the corresponding fraction.The fractions $\rho_j(t)$ are normalized:
\begin{equation}
\sum_j \rho_j(t)\ =1\;,\qquad \forall t\ .
\end{equation}

Equation (\ref{pripasex}) yields the 
average number of offspring of a single
individual. Assuming that reproduction events are 
independent from one another, the number of offspring
of the single individual is Poisson distributed with 
average equal to the expected value 
$\overline{\nu(\alpha)}$ (eq.~(\ref{pripasex})).
As a consequence, the number of 
offspring of a whole deme $\delta$ is also Poisson 
distributed, with average
$Q(\delta)$ given by the sum of $\overline{\nu(\alpha)}$ over its members:
\begin{equation}
Q(\delta)=\sum_{\alpha\in\delta}\overline{\nu(\alpha)}\ .
\end{equation}
It follows from Eq.~(\ref{pripasex}) that the size of the offspring
of the $j$-deme $\delta$ depends only on the number $j$ of its altruists and 
is given, on average, by $Q(\delta)=V_jL$, where
\begin{equation}
V_j=\frac{T_j}{\sum_{k=0}^L T_k\rho_k(t)}\ .
\end{equation}
In this equation we have introduced the function
\begin{equation}
T_j=\frac{1-rj/L}{1-c\,\theta(j-j^*)}\ ,
\end{equation}
where $\theta(x)$ is the unit step function.
We stress that this is the {\em average\/} new size of the deme, 
disregarding its new genetic composition. Indeed,
the genetic composition of a new deme which originates from a
$j$-deme is also random.
We recall that the number of altruists in the
old deme is $j$ while the corresponding number of egoists is $L-j$.
We also recall that the relative probability 
of survival in the same deme is $1-r$.
The probability that
a given offspring of a $j$-deme is altruist is therefore
\begin{equation}
p_j=\frac{(1-r)\,j}{L-rj}\ .
\end{equation}
Since reproduction events are independent,
the number of altruists
is binomially distributed with success probability $p=p_j$.
This means that the average fraction of altruists is equal to $p_j$, while
its variance is approximately given by
\begin{equation}
\sigma_j^{\rm rep}=p_j(1-p_j)\,\frac{1}{V_jL}\ .
\end{equation}
This expression holds in the approximation in which the average
size of the offspring deme is substituted for its actual size:
due to the reproduction, demes grow up and shrink, so that
their actual size is not constant.

We now consider the effects of deme splitting.
It produces new demes with 
smaller size but with the same composition in average.
Assuming independence, we have again a binomial redistribution of
individuals in the splitted demes.
The average size is brought back to $L$ and the average fraction of
altruists is still given by $p_j$,
while its variance is given by
\begin{equation}
\sigma_j^{\rm split}=\ p_j\ (1-p_j)\frac{1}{L}\ .
\end{equation}

If the deme size $L$ is large enough, we can approximate
the distribution of altruist numbers in the new demes by
a gaussian. Define
\begin{equation}
G_{ij}=\exp\left(-\frac{(i-p_jL)^2}{2\sigma_jL^2}\right)\;,
\qquad (j\neq 0,L)\ ,
\end{equation}
where 
\begin{equation} \label{sigma}
\sigma_j= \sigma_j^{\rm split}+\sigma_j^{\rm rep}\ .
\end{equation}
The transition probability matrix
(i.e., the probability of having $i$ altruist in
a deme at time $t+1$ from a $j$-deme at time $t$,
after the reproduction and splitting steps) is given by
\begin{equation}
A_{ij}=\frac{G_{ij}}{\sum_i G_{ij}}\;,
\qquad (j\neq 0,L)\;.
\end{equation}
Since, in the absence of mutations,
all demes originating
from a homogeneous deme (be it altruistic or egoistic)
are also homogeneous and of the same type, we have
\begin{eqnarray}\label{nomut}
A_{LL}&=&A_{00}=1;\\
A_{iL}&=&0,\qquad i\neq L;\\
A_{i0}&=&0,\qquad i\neq 0
\end{eqnarray}
It should be remarked that only large demes split and therefore 
$\sigma_j=\sigma_j^{\rm rep}\ $ for small demes. Nevertheless the error we
introduce extending (\ref{sigma}) to all demes is negligible, because for small
demes $V_j < 1$, so that $\sigma_j^{\rm split} < \sigma_j^{\rm rep}\ .$

The number of demes changes because of splitting:
however, since the number of demes, $N$, goes to infinity,
the fraction of demes originating from $j$-demes is equal to its average
because of the law of large numbers. This number is 
given by 
\begin{equation}
V_j \rho_j(t)=\frac{T_j\rho_j(t)}{\sum_{k=0}^L T_j\rho_j(t)}\ ,
\end{equation}
We are thus led to the equation
\begin{equation}
\rho_i(t+1)= \frac{1}{{\cal Z}(t)}\sum_j  A_{ij} T_j\rho_j(t)\ ,\label{qs}
\end{equation}
which closely resembles Eigen's quasispecies equation\cite{Eig}.
The normalization factor ${\cal Z}_t$ is given by
\begin{equation}
{\cal Z}(t)=\sum_j T_j\rho_j(t)
\end{equation}
The main difference between eq.~(\ref{qs}) and the quasispecies equation
lies in the fact that the matrix of transition probabilities $(A_{ij})$
also depends on the vector $(\rho_j(t))$ via the variances $(\sigma_j)$.

\section{Results}
In the long run, the population will reach a steady
distribution, which may be of one among three
types.
The first two types correspond to a population  
composed only by altruistic or egoistic individuals.
It is trivial to
remark that the demes for these two phases have an homogeneous composition.
The third type correspond to a situation where, together with
homogeneous demes of both kinds, one has inhomogeneous demes,
formed by both egoistic and altruistic individuals.

Let us first imagine that the population is initially composed
only by homogeneous demes of either kind: completely
altruistic and completely egoistic.
In this case no mixed demes can be generated,
since homogeneous demes only reproduce themselves.
The advantage for the completely altruistic deme is
$1/(1-c)$ while its disadvantage is $1-r$.
It is easy to understand that, in this case,
the population will end up being made of
only completely altruistic demes if
$c>r$ and only egoistic ones if $c<r$.
In other words:
\begin{mathletters}
\begin{eqnarray} 
\rho_{\rm eq}(j)&=&\cases{1, &for $j=0$;\cr
0, &for $j>0$;}\qquad {\rm for\ }c<r;\\
\rho_{\rm eq}(j)&=&\cases{0, &for $j<L$;\cr
1, &for $j=L$;}\qquad {\rm for\ } c>r.
\end{eqnarray}
\end{mathletters}
If we mark with dots the region in which only altruistic
individuals are present and with short vertical lines
the region where only egoistic ones are present in the $(r,c)$ plane,
we obtain the ``phase diagram'' of Fig.~1. We call ``phases''
the different regimes of the equilibrium population: egoistic,
altruistic, or mixed.
\begin{figure}
\centerline{\epsfxsize=10cm
\epsffile{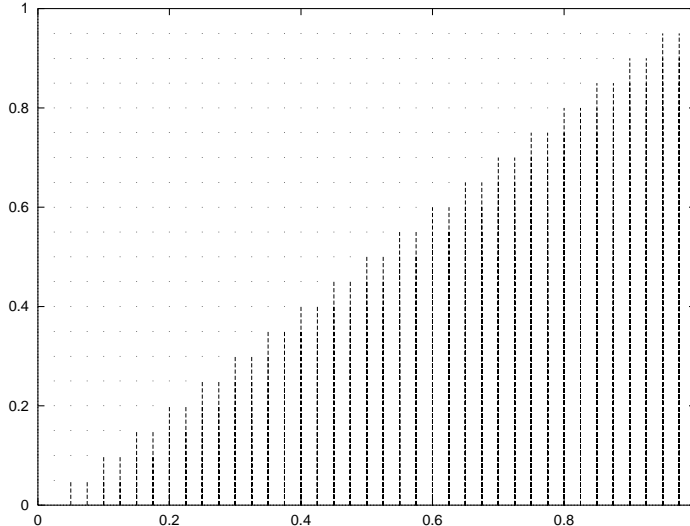}
}
\caption{Phase diagram in the ($r,c$) space for $L$
(deme size) equal to 20 and $j^*$ (threshold) equal to 4. 
One observes the stability of a wholly altruistic 
population (dots) for $c>r$ and a fully egoistic 
one (vertical lines) for $c<r$.}
\label{fig1}
\end{figure}

Suppose now that also inhomogeneous demes are initially present.
In this case that demes with
$j\ge j^*$ share with completely altruistic ones
the ``altruistic advantage'' and, since they have
$L-j$ egoistic individuals, they partially share 
the ``egoistic advantage'' with the completely
egoistic demes.
From this point of view, the fittest demes would be those with $j=j^*$.
Nevertheless, reproduction and splitting
changes their composition, so that they
most of the times do not reproduce.
Briefly, mixed demes can generate homogeneous demes but 
not viceversa.
Due to these competing effects,
it not {\em a priori\/} clear
if the dynamics may lead  to mixed populations.
If it is so, we expect that this will happen provided that $j^*$
is so small that the mixed 
demes can take advantage of both altruistic and egoistic
behavior in the best way.
This is shown by the equilibrium solution of the master equation.
Compare, in fact, Fig.~1 with Fig.~2, in which $L=20$ again,
but $j^*=1$ instead of 4.
One sees the co-existence, at equilibrium,
of both altruistic and egoistic individuals.
\begin{figure}
\centerline{
\epsfxsize=10cm
\epsffile{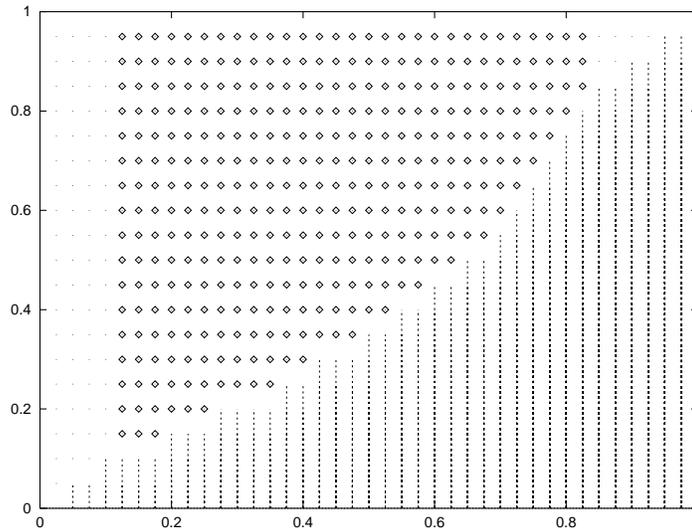}
}
\caption{Phase diagram in the ($r,c$) space for $L=20$ and $j^*=1$. 
One observes the appearance of a region of stability of an
{\em inhomogeneous\/} (mixed altruistic-egoistic) population
(diamonds).}
\label{fig2}
\end{figure}

Figs.~3 and 4 correspond to intermediate situations with
$j^*=2$ and $j^*=3$.
It is clear that the region where mixed populations
are stable becomes progressively smaller from $j^*=1$ to
$j^*=3$ and disappears for $j^*=4$.     
\begin{figure}
\centerline{
\epsfxsize=10cm
\epsffile{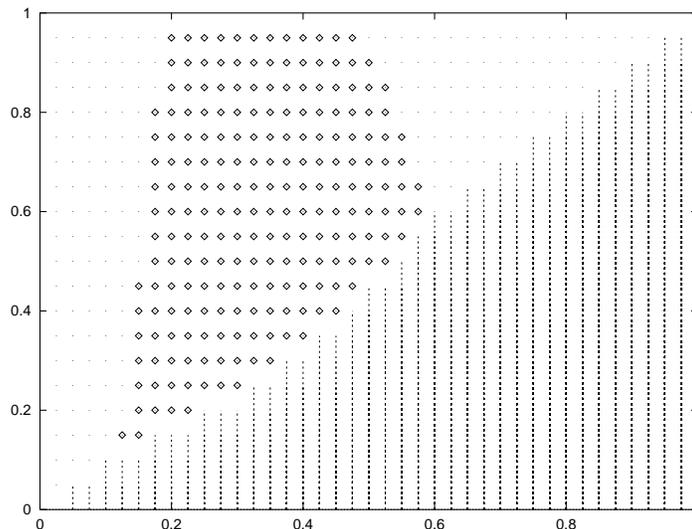}
}
\caption{Phase diagram in the ($r,c$) space for $L=20$ and $j^*=2$. 
The region of stability of the inhomogeneous population (diamonds)
has shrinked.}
\label{fig3}
\end{figure}

\begin{figure}
\centerline{
\epsfxsize=10cm
\epsffile{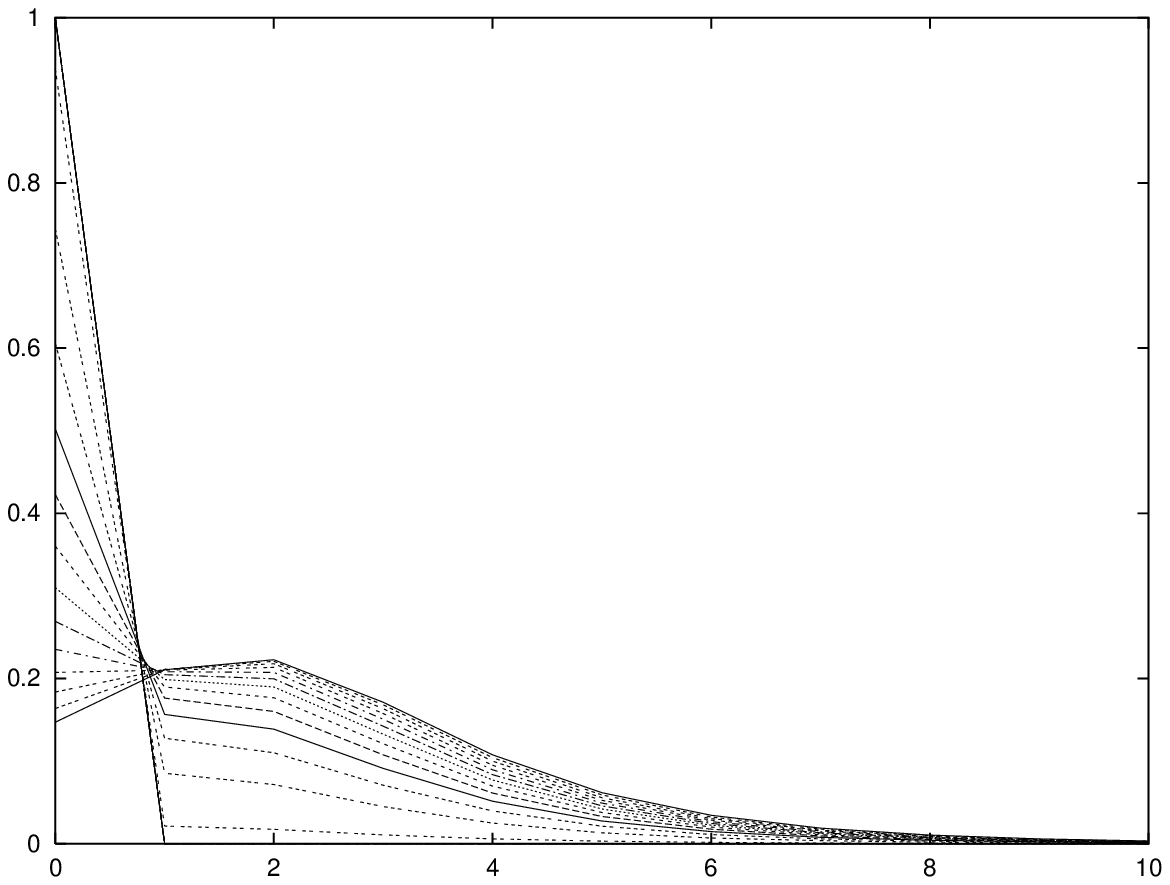}
}
\caption{Distribution of the number $j$ of individuals
in each deme for $L=20$, $j^*=2$, $r=0.4$, 
$c=0.0,\; 0.05,\; 0.1,\; \dots,\; 0.95$. One sees that
most demes are inhomogeneous ($j\neq 0,L$) and none is
formed only by altruistic individuals.}
\label{fig5}
\end{figure}

\begin{figure}
\centerline{
\epsfxsize=10cm
\epsffile{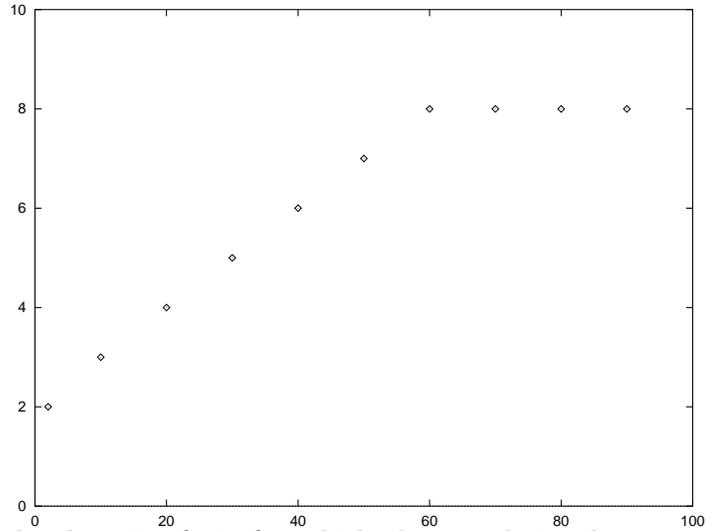}
}
\caption{The critical value $j^{*}_{cr}$ of $j^*$, for which the
population becomes homogeneous,
as a function of deme size $L$. For $L<60$, one has $j_{\rm cr}^*\propto L$,
while $j_{\rm cr}^*$ attains a constant value for $L>60$.}
\label{fig6}
\end{figure}

Let us stress that the equilibrium distribution
$\rho_{\rm eq}(j)$ is non-trivial when the equilibrium
population is inhomogeneous (see, for example, Fig.~\ref{fig5}).
In fact, together with mixed demes,
also homogeneous demes are present since they
arise as offsprings of mixed demes at any generation.
The edges of the stability region of
the mixed phase depend both on $L$ and on $j^*$.
Figures 1--4 correspond
to $L=20$, but for larger $L$ the qualitative behaviour it is the same:
if we increase $j^*$ the domain of mixed phase becomes smaller and, above a 
critical value ($j^{*}_{cr}$), it desappears. 
The value of $j^{*}_{cr}$ depends on $L$.
Plotting $j^{*}_{cr}$ vs.\ $L$, we obtain the graph 
represented in Fig.~\ref{fig6}, from which one clearly sees that 
\begin{equation}
\lim_{L \rightarrow \infty} j^{*}_{cr}/L\, = 0\, ,
\end{equation}
while $j^{*}_{cr}/L\, ={\rm const.}$, for $L<60$.
This behaviour suggests that for large demes 
the mixed phase can exist only
if the altruistic advantage is granted by a small
number of altruistic individuals.

\section{Discussion}
The most important feature of our model is that {\em selection acts at 
individual level}, although 
the individual survival probability takes into account 
also of the ``interaction" among individuals. This very simple interaction is 
modeled by the dependence of the fitness function (see Tab.~1) on 
the genetic composition of group:
the individual survival probability is sensitive to the presence of 
other altruists in the group. So, without invoking either a kin selection
mechanism or a group selection one for the 
maintenance of altruists in the population, we 
have seen that, when the group 
advantage balances the individual (altruistic) disadvantage, mixed groups may 
co-exist at equilibrium. This kind of equilibrium also depends on the 
percentage of altruists in a deme (for small enough demes, i.e., $L<60$); for 
larger demes a finite number of altruists is sufficient to produce
inhomogeneous demes at equilibrium.

Let us remark that the stabilization of inhomogeneous populations
is reached in the absence of the (rather improbable) mutations
of the ``altruistic'' into the ``egoistic'' gene (and viceversa):
the key role is played by the fluctuations in the composition
of daughter demes from inhomogeneous ones. These fluctuations are
even more important when, as in Ref.~\cite{D}, the size of the
daughter demes is allowed to vary.

Summarizing, we have shown in a numerically solvable model that
if the advantage of the altruistic behavior of a few individuals
falls (above a given threshold)
on the whole of a sufficiently large deme, a stable situation
arises in which inhomogeneous demes (consisting of both altruistic
and egoistic individuals) coexist with homeogeneous ones.
For large average deme size, the average number of altruistic
individuals approaches the threshold.

\section*{Acknowledgements}
We thank Dr.~Ugo Bastolla for enlightening discussions. 
R.D. thanks in particular Dr. Erik Aurell for encouragement
and the Mathematical Department of the University
of Stockholm for financial support.

\end{document}